\documentclass[12pt,a4paper]{article}
\usepackage{graphicx}
\usepackage[latin2]{inputenc}
\usepackage{latexsym}
\usepackage{epsf}
\usepackage{epsfig}
\usepackage{cite}

\def\bea{\begin{eqnarray}}  
\def\eea{\end{eqnarray}}
\def\bc{\begin{center}}
\def\ec{\end{center}}

\def\lrvec#1{\vbox{\ialign{##\crcr
${\hspace{1pt}\scriptscriptstyle\leftrightarrow\hspace{-1pt}}
$\crcr\noalign{\nointerlineskip}
$\hfil\displaystyle{#1}\hfil$\crcr}}}

\begin{document}
\pagestyle{empty}
\begin{flushright}
\end{flushright}
\vspace*{5mm}
\begin{center}

{\large {\bf Gauge invariance of the $h^0\rightarrow\gamma\gamma$ amplitude}}\\
\vspace*{1cm}

{\bf Piotr~H.~Chankowski}\footnote{Email:chank@fuw.edu.pl} and 
{\bf Adrian Lewandowski}\footnote{Email:adrian.lewandowski@fuw.edu.pl},\\
\vspace{0.5cm}
 
Faculty Physics, \\
University of Warsaw , Ho\.za 69, 00-681, Warsaw, Poland

\vspace*{1.7cm}
{\bf Abstract}
\end{center}
\vspace*{5mm}
\noindent
{We point out that the one-loop amplitude of the $h^0\rightarrow\gamma\gamma$ 
decay is gauge invariant owing to a particular relation between the 
trilinear couplings and the Higgs boson mass. This relation follows 
only from the gauge symmetry breaking pattern realized by the potential 
of the scalar fields and not on its specific form. This allows to justify
the seemingly inconsistent calculation of the $h^0\rightarrow\gamma\gamma$
amplitude in the minimal supersymmetric model (MSSM) in which one takes the 
mass of the lighter Higgs boson from e.g. the one-loop effective potential. 
}
\vspace*{1.0cm}
\date{\today}


\vspace*{0.2cm}
 
\vfill\eject
\newpage

\setcounter{page}{1}
\pagestyle{plain}

The decay into two photons was one of the most important channels in 
which a spin 0 particle of mass around 125 GeV was discovered at the LHC 
one year ago \cite{LHCdisc}. While the measured decay rates and other 
characteristics of the discovered particle agree well with the properties 
of the Higgs boson predicted by the minimal version of the standard theory, 
the Standard Model (SM), they do not preclude the possibility that it is the 
first discovered particle of an extended Higgs sector consisting of more 
spin 0 states, like e.g. that of the two-doublet models or of the minimal 
supersymmetric model (MSSM). Despite the lack of any specific signal which 
would favour the supersymmetric interpretation, the latter possibility 
still appears attractive to many, because of the hierarchy problem.

Yet the calculation of the $h^0\rightarrow\gamma\gamma$ amplitude in 
the MSSM, unlike the seemingly analogous calculation within the SM or 
two-doublet models, has a subtle point which is usually overlooked,
because, assuming gauge invariance from the beginning, one usually 
computes only one of the relevant formfactors, without checking the 
others. In this note we give a simple formula for the $W^\pm$ bosons 
contribution to the $h^0\rightarrow\gamma\gamma$ amplitude valid for
any scalar sector breaking the $SU(2)_L\times U(1)_Y$ symmetry down to
$U(1)_{\rm EM}$ and using it we give a justification of the standard way of 
calculating the MSSM prediction for the $h^0\rightarrow\gamma\gamma$ rate.
\vskip0.2cm

The general form of the $h^0\gamma\gamma$ vertex is
\begin{eqnarray}
-i{\cal A}^{\mu\nu}=-i{e^2\over16\pi^2}\left(F_1g^{\mu\nu}+F_2s^\mu s^\nu
+F_3l^\mu l^\nu+F_4l^\mu s^\nu+F_5s^\mu l^\nu\right),
\end{eqnarray}
with $F_i\equiv F_i(l,s)$. By the Bose symmetry $F_i(l,s)=F_i(s,l)$ for 
$i=1,4,5$ and $F_2(l,s)=F_3(s,l)$.
Because of the non-Abelian origin of the electromagnetic $U(1)$ group, gauge 
invariance requires only \cite{PESIO} that 
$l_\nu{\cal A}^{\mu\nu}\epsilon^\ast_\mu(s)=0$
and $s_\mu{\cal A}^{\mu\nu}\epsilon^\ast_\nu(l)=0$ for $(l+s)^2\equiv q^2=M^2_h$.
Thus, because $k\!\cdot\!\epsilon^\ast(k)=0$, the formactor $F_5$ is 
unconstrained by gauge invariance while the remaining ones must satisfy
\begin{eqnarray}
F_1+s^2F_2+(l\!\cdot\!s)F_4=0\phantom{aaaa}{\rm for}\phantom{aa}
2(l\!\cdot\!s)=M_h^2-s^2~\!,\phantom{aa}l^2=0~\!,\nonumber\\
F_1+l^2F_3+(l\!\cdot\!s)F_4=0\phantom{aaaa}{\rm for}\phantom{aa}
2(l\!\cdot\!s)=M_h^2-l^2~\!,\phantom{aa}s^2=0~\!.\nonumber
\end{eqnarray}
On shell, i.e. for $l^2=s^2=0$, one must of course have 
$F_1=-(l\!\cdot\!s)F_4$ and the amplitude reduces to the standard form
\begin{eqnarray}
{\cal M}={e^2\over16\pi^2}~\!{F_1\over(l\!\cdot\!s)}~\!
[(l\!\cdot\!s)g^{\mu\nu}-l^\mu  s^\nu]\epsilon^\ast_\mu(s)\epsilon^\ast_\nu(l)~\!,
\end{eqnarray}
but contrary to the naive expectation the formfactors $F_2$, $F_3$ 
and $F_5$ do not 
vanish even on-shell.\footnote{Off-shell the formfactors $F_2$, $F_3$ and 
$F_5$ do not vanish even in an Abelian theory (e.g. if the neutral particle 
$h^0$ couples to photons only by loops of a spin zero charged particle) in 
which gauge invariance imposes a stronger constraint 
$l_\nu{\cal A}^{\mu\nu}=s_\mu{\cal A}^{\mu\nu}=0$ for arbitrary $l$ and $s$, 
because for off-shell photons there are two gauge 
invariant structures: $[(l\!\cdot\!s)g^{\mu\nu}-l^\mu  s^\nu]F_1/(l\!\cdot\!s)$
and $[s^2l^2 s^\nu l^\mu-l^2(l\!\cdot\!s)s^\mu s^\nu-s^2(l\!\cdot\!s)l^\mu l^\nu
+(l\!\cdot\!s)^2s^\mu l^\nu]F_5/(l\!\cdot\!s)^2$. On-shell one then has
$F_2=F_3=0$ but still $F_5\neq0$.}

As will become clear, the relation $F_1+(l\!\cdot\!s)F_4=0$ is
satisfied only if the ``kinematical'' mass squared
$(M^2_h)_{\rm kin}\equiv(l+s)^2=2(l\!\cdot\!s)$ is related to a trilinear 
coupling of $h^0$ to the would-be Nambu-Goldstone bosons. Therefore 
using in the MSSM calculation 
$(M_h)_{\rm kin}=125$ GeV (that is, implicitly including in $M_h$ higher loop
corrections, as is necessary in the MSSM to overcome the tree level 
bound $M^2_h\leq M^2_Z$) 
potentially leads to violation of gauge invariance.
We will argue however, that a gauge invariant result can be obtained
if one derives both the Higgs boson mass {\sl and} the trilinear
couplings from the same effective potential (which includes the most 
important corrections to the Higgs boson mass).
\vskip0.2cm

We consider a general set of real scalar fields $\phi_i$ with a potential 
$V(\phi)$. On the scalar fields the $SU(2)_L\times U(1)_Y$ symmetry of 
the electroweak interactions is realized through the purely imaginary, 
antisymmetric generators $T^\pm\equiv T^1\pm iT^2$, $T^0\equiv T^3-Y$ 
and $Q\equiv T^3+Y$ of which the first three are broken by the VEV 
$\langle\phi_i\rangle=v_i$ (that is, $T^{0\pm}v\neq0$ while $Qv=0$). 
The matrix ${\cal M}^2_{ij}\equiv\left.V^{(2)}_{ij}\right|_v$ (we use the 
notation $V^{(n)}_{ij\dots}\equiv\partial^nV/\partial\phi_i\partial\phi_j\dots$) 
of the masses squared of the spin zero fields has therefore three zero 
eigenvalues with the zero eigenvectors $u^i_{(0)}\propto T^0_{ij}v_j$ and
\begin{eqnarray}
u^i_{(\pm)}=N_{(\pm)}T^\pm_{ij}v_j~\!,\phantom{aaaa}
N_{(+)}=-N_{(-)}={1\over\sqrt{v~\!T^-T^+v}}={1\over\sqrt{v~\!T^+T^-v}}~\!.
\nonumber
\end{eqnarray}
where the normalization factors are such that the eigenvectors $u^i_{(\pm)}$
are orthonormal: $(u^i_{(\pm)})^\ast u^i_{(\pm)}=u^i_{(\mp)} u^i_{(\pm)}=1$
(the signs are chosen in agreement with the usual SM convention).
Since the mass squared of the $W^\pm$ bosons is given by
\begin{eqnarray}
M^2_W={1\over4}~\!g^2_2~\!v(T^-T^++T^+T^-)v
={1\over2}~\!g^2_2\left(v~\!T^+T^-v\right)\equiv{1\over4}~\!g^2_2v^2_H,
\end{eqnarray}
(where $v_H=246$ GeV) one has
\begin{eqnarray}
u^i_{(\pm)}=\pm {g_2\over\sqrt2M_W}~\!T^\pm_{ij}v_j~\!.
\end{eqnarray}
Furthermore,  we assume that ${\cal M}^2_{ij}$ has, among others, an
eigenvector $u^i_{(h)}$ corresponding to a neutral spin zero particle $h^0$
of mass $M_h$: ${\cal M}^2_{ij}u^j_{(h)}=M^2_hu^i_{(h)}$, so that 
\begin{eqnarray}
\phi_i=u^i_{(+)}G^++u^i_{(-)}G^-+u^i_{(h)}h^0+\dots ,
\end{eqnarray}
where $G^\pm$ are the would-be Nambu-Goldstone boson fields and
the ellipses stand for other terms, irrelevant for our analysis. 

Differentiating twice the usual symmetry condition 
$V^{(1)}_i(\phi)T^\pm_{ij}\phi_j=0$ satisfied by the potential and contracting 
the result from the right with $T^\mp v$ one gets, setting $\phi_i=v_i$, 
the relation:
\begin{eqnarray}
V^{(3)}_{lki}T^\pm_{ij}v_j~\!T^\mp_{km}v_m+V^{(2)}_{ki}T^\pm_{il}T^\mp_{km}v_m
+V^{(2)}_{li}T^\pm_{ik}T^\mp_{km}v_m=0~\!.\nonumber
\end{eqnarray}
The second term vanishes ($T^\mp_{km}v_m$ is the zero eigenvector of the
masses squared matrix $V^{(2)}_{ij}$) and the remaining two projected onto 
the direction of the eigenvector $u^i_{(h)}$ give the crucial relation 
\begin{eqnarray}
\kappa={g^2_2\over2M^2_W}~\!u^j_{(h)}{\cal M}^2_{ji}(T^\pm T^\mp)_{ik}v_k
={g^2_2\over2M^2_W}~\!M^2_h~\!(u_{(h)}T^\pm T^\mp v)~\!,\label{eqn:main}
\end{eqnarray}
between the mass $M_h$ and the trilinear coupling $\kappa$ 
defined by
\begin{eqnarray}
{\cal L}=-\kappa~\!
h^0G^+G^-+\dots\equiv-V^{(3)}_{lki}u^l_{(+)}u^k_{(-)}u^i_{(h)}~\!h^0G^+G^-
+\dots\label{eqn:kappadef}
\end{eqnarray}
(from (\ref{eqn:main}) it follows that $u_{(h)}T^+T^-v=u_{(h)}T^-T^+v$).
Within the same setting it is also easy to get the following terms
of the Lagrangian relevant for calculating the $h^0\rightarrow\gamma\gamma$
amplitude
\begin{eqnarray}
{\cal L}=-{ig^2_2\over2M_W}~\!(u_{(h)}T^+T^-v)\left(
W^+_\mu(G^-\lrvec\partial_\mu h^0)-W^-_\mu(G^+\lrvec\partial_\mu h^0)
\right)\nonumber\\
+eM_WA_\mu(W^+_\mu G^-+W^-_\mu G^+)\phantom{aaaaaaaaaaaaaaaaaaa}
\nonumber\\
-{eg^2_2\over2M_W}~\!(u_{(h)}T^+T^-v)~\!A^\mu(W^+_\mu G^-+W^-_\mu G^+)h^0
\phantom{aaaaa}~\!\nonumber\\
+g^2_2~\!(u_{(h)}T^+T^-v)~\!W^+_\mu W^-_\mu h^0
\phantom{aaaaaaaaaaaaaaaaaaa}~\nonumber\\
+{1\over2}~\!\xi g^2_2~\!(u_{(h)}T^+T^-v)~\!(\bar\eta_+\eta_++\bar\eta_-\eta_-
)h^0~\!,\phantom{aaaaaaaaaa}
\end{eqnarray}
where $\xi$ is the gauge parameter and $\eta_\pm$, $\bar\eta_\pm$
are the relevant ghost fields.
\vskip0.2cm

The one-loop Feynman diagrams in the 't Hooft gauge $\xi=1$
contributing to the $h^0\rightarrow\gamma\gamma$ amplitude can 
be easily expressed in terms of the $C_0$ and $C_{ij}$ three-point 
functions defined in \cite{AXELROD}. In turn, the $C_{ij}$ functions 
can, for $l^2=s^2=0$, and all masses circulating in loops equal to $M_W$,
be reduced to the scalar $C_0$ function and the two-point  functions
$b_0(0,M_1,M_2)$ or $b_0(q^2,M_1,M_2)$. In the formfactors $F_1$, $F_2$,
$F_3$ and $F_4$ all $b_0$ functions cancel out and for $F_4$ and $F_1$ 
one obtains the following results\footnote{We have explicitly
checked that indeed, $F_2=F_3\neq0$ and $F_5\neq0$.}
\begin{eqnarray}
F_4=g_2^2~\!(u_{(h)}T^+T^-v)~\!{1\over l\!\cdot\!s}\left\{
{M^2_h\over M^2_W}\left(1-2M^2_WC_0\right)+6-12M^2_WC_0+16(l\!\cdot\!s)C_0
\right\},\label{eqn:F4}
\end{eqnarray}
\begin{eqnarray}
F_1=g_2^2~\!(u_{(h)}T^+T^-v)\left\{
-{M^2_h\over M^2_W}\left(1-M^2_WC_0\right)-6+12M^2_WC_0-14(l\!\cdot\!s)C_0
\right\},\label{eqn:F1}
\end{eqnarray}
Only if the ``kinematical'' mass squared $2(l\!\cdot\!s)$ equals the 
parameter $M^2_h$ arising from the coupling (\ref{eqn:main})
is the relation $F_1=-(l\!\cdot\!s)F_4$ imposed by gauge invariance
satisfied. In this case one gets
\begin{eqnarray}
F_4={1\over l\!\cdot\!s}\left[{2\over v_H}~\!(u_{(h)}T^+T^-v)\right]
\left\{{M^2_h\over v_H}\left(2+12M^2_WC_0\right)+
{M^2_W\over v_H}\left(12-24M^2_WC_0\right)\right\},
\end{eqnarray}
which, for the Higgs sector of the SM, for which the expression in the
first bracket equals 1, after taking into account that 
\begin{eqnarray}
{1\over2}~\!M^2_h~\!C_0=f(\tau)=
\left\{
\matrix{\phantom{aa}{\rm arcsin}^2(1/\sqrt\tau)\phantom{aaaaaa}
{\rm for}\phantom{aa}\tau\geq1\cr
-{1\over4}\left(\ln{1+\sqrt{1-\tau}\over1-\sqrt{1-\tau}}-i\pi\right)^2
\phantom{aa}{\rm for}\phantom{aa}\tau<1}
\right.
\end{eqnarray}
where $\tau=4M_W^2/M^2_h$, gives the well known result
\cite{ELGANA,SHVAVOZA}:
\begin{eqnarray}
{\cal M}=-{e^2\over8\pi^2v_H}\left[2+3\tau+3\tau(2-\tau)f(\tau)\right]
[(l\!\cdot\!s)g^{\mu\nu}-l^\mu s^\nu]\epsilon_\mu^\ast(s)\epsilon_\nu^\ast(l)~\!.
\label{eqn:Amplitudefinal}
\end{eqnarray}
In the general case the contribution of the gauge bosons to the 
$h^0\rightarrow\gamma\gamma$ amplitude is obtained with the help of the
replacement
\begin{eqnarray}
{1\over v_H}\rightarrow {g^2_2\over2M_W^2}~\!(u_{(h)}T^+T^-v)~\!.
\end{eqnarray}
Performing symbolic calculations with the help of the package FEYNCALC 
\cite{FEYNCALC} we have verified that one gets the same result 
(\ref{eqn:Amplitudefinal}) for arbitrary value of the gauge fixing
parameter $\xi$ provided $(M^2_h)_{\rm kin}=2(l\!\cdot\!s)=M^2_h$.
\vskip0.2cm

In models like the Standard Model or its various nonsupersymmetric two 
Higgs doublet extensions, in which the mass $M_h$ can be taken for a free 
adjustable parameter, gauge invariance requirements are obviously satisfied.
However, in the MSSM, if the amplitude is computed using the standard
Feynman rules (derived from the fundamental Lagrangian) one cannot freely 
set $2(l\!\cdot\!s)=(125~{\rm GeV})^2$ in (\ref{eqn:F4}) and (\ref{eqn:F1})
because the parameter $M^2_h$ arising from the trilinear $h^0G^+G^-$
coupling is bounded by $M_Z^2$. Fortunately the way we have derived
the amplitude shows that the gauge invariance is ensured if the 
necessary coupling $\kappa$  in (\ref{eqn:kappadef}) and the Higgs 
mass are computed from the one-loop effective potential satisfying the
same symmetry condition as the one we have used to derive the relation
(\ref{eqn:main}). Renormalizability of the potential used is not 
necessary for finitness of the amplitude. Moreover, owing to the fact 
that the one-loop $h^0\rightarrow\gamma\gamma$ amplitude is the same for 
all values of the gauge parameter $\xi$, the vertex factor $\kappa$  
can safely be taken from the full one-loop effective potential 
including also contributions of gauge bosons sector which is most 
easily computed in the Landau gauge. 
\vskip0.2cm

The whole point of this note was to lend some justification to the usual 
practice of computing the decay rate of the supersymmetric lightest neutral 
CP-even Higgs boson into two photons by taking for the gauge/Higgs
sector contribution the SM model result (\ref{eqn:Amplitudefinal}) 
modified by cosine of the Higgs neutral Higgs boson mixing angle. 
Since the effective  potential of the MSSM is able to incorporate
the dominant contributions allowing to lift the mass of the 
lightest neutral CP-even Higg boson $h^0$ to 125 GeV
(see e.g. \cite{CAQUWA}), this solves the 
problem of a consistent calculation of the $h^0\rightarrow\gamma\gamma$ 
amplitude. In this approach the factor 
$(u_{(h)}T^+T^-v)$ equals obviously ${1\over2}v_H\cos\alpha$ 
where $\alpha$ is the mixing angle diagonalizing the $h^0$ and $H^0$ mass 
squared matrix derived from the effective potential.

\vskip1.0cm
\section*{Acknowledgments}
P.H.Ch. wants to thank Janusz Rosiek, with whom he
discovered the problem many years ago, for discussion and interest.

\end{document}